\documentclass[10pt]{article}

\usepackage{epsfig}

\def\lsim{\mathrel{\rlap{\lower4pt\hbox{\hskip1pt$\sim$}}
    \raise1pt\hbox{$<$}}}         
\def\gsim{\mathrel{\rlap{\lower4pt\hbox{\hskip1pt$\sim$}}
    \raise1pt\hbox{$>$}}}         

\def\overleftrightarrow#1{\vbox{\ialign{##\crcr
    $\leftrightarrow$\crcr
    \noalign{\kern 1pt\nointerlineskip}
    $\hfil\displaystyle{#1}\hfil$\crcr}}}

\newcommand{\be}{\begin{equation}}
\newcommand{\ee}{\end{equation}}
\newcommand{\bea}{\begin{eqnarray}}
\newcommand{\eea}{\end{eqnarray}}

\begin{document}

\hfill {\bf TUM/T39-03-26} \\
\vspace{0.01in}
\hfill {October 2003} \\

\vspace{0.25in}

\begin{center}
{\bf \Large Quasiparticle Description of Hot QCD\\at Finite Quark
Chemical Potential\footnote{Work supported in part by BMBF and GSI}}
\end{center}
\vspace{0.25 in}
\begin{center}
{M.A. Thaler$^a$, R.A. Schneider$^{a,b}$, W. Weise$^{a,b}$}\\
\vspace{0.4cm}
{\small \em  $^a$Physik-Department, Technische Universit\"{a}t M\"{u}nchen
D-85747 Garching,
GERMANY\\}

{\small \em  $^b$ECT$^*$, I-38050 Villazzano (Trento), ITALY\\}
%
\end{center}
\vspace{0.25 in}
\begin{abstract}
We study the extension of a phenomenologically successful
quasiparticle model that describes lattice results of the equation of
state of the deconfined  phase of QCD for $T_c\le T\lsim 4 T_c$, to
finite quark chemical potential $\mu$. The phase boundary line $T_c(\mu)$, the
pressure difference $\Delta p(T,\mu)=(p(T,\mu)-p(T,\mu=0))/T^4$ and
the quark number density $n_q(T,\mu)/T^3$ are calculated and compared
to recent lattice results. Good agreement is found up to quark
chemical potentials of order $\mu\sim T_c$.

\end{abstract}
\newpage
\section{Introduction}
The phase structure of QCD at high temperature and non-vanishing
baryon chemical potential has been subject of intense research in
recent years. Heavy-ion collisions at high energies have been and are
being explored at SPS/CERN and RHIC/BNL \cite{Heinz:2002gs} in search
for signals of the Quark-Gluon Plasma (QGP). Large-scale lattice QCD
computations at finite temperature have been performed
\cite{Gottlieb:1996ae,Karsch:2000ps,AliKhan:2001ek}, and first
extensions to non-zero baryon chemical  potential appear now to be
feasible. It has proven possible to trace out the phase boundary line
$T_c(\mu)$ separating the hadronic phase from the QGP phase
for $N_f=4$ \cite{Fodor:2001au,D'Elia:2002gd}, $N_f=2$
\cite{deForcrand:2002ci} and $N_f=3$ \cite{deForcrand:2003hx} flavors
of quarks up to quark chemical potentials $\mu$ of order $T_c$. First
numerical results for the QCD Equation of State (EoS), i.e. the
pressure $p(T,\mu)$ and the quark number density $n_q(T,\mu)$ are also
available for $N_f=2+1$ \cite{Fodor:2002km} and $N_f=2$ \cite{Allton:2003vx}.
As well as being of intrinsic theoretical interest,
such studies provide conceptual guidance for current heavy ion collision
experiments at SPS and RHIC, where the chemical freeze-out occurs at
$\mu_{f.o.}\simeq 100$ MeV, (baryon chemical potential
$\mu_B\simeq 300$ MeV) \cite{Braun-Munzinger:1999qy} and
$\mu_{f.o.}\simeq 15$ MeV, ($\mu_B\simeq
45$ MeV) \cite{Braun-Munzinger:2001ip},
respectively.\\[0.5cm]
Systematic perturbative expansions of the QCD equation of state within
the framework of thermal field theory show bad convergence even for
very large temperatures (several times $T_c$) far beyond the region
accessible to present experiments \cite{Kajantie:2002wa}. 
Various techniques, such as dimensional reduction, screened
perturbation theory or hard-thermal loop (HTL) perturbation theory show
better convergence and good agreement with lattice results for $T\gsim
3 T_c$ \cite{Blaizot:2003tw}. Various interpretations of the lattice
data have been attempted in terms of physical quantities, most
prominently as the EoS of a gas of massive quark and
gluon quasiparticles. Their thermally generated masses are based on
perturbative calculations carried out in the HTL scheme
\cite{Peshier:1995ty,Levai:1997yx,Peshier:1999ww}. This approach has been
extended to non-vanishing quark chemical potential and good agreement
with finite $\mu$ lattice calculations for $N_f=2+1$ flavors has been 
found \cite{Szabo:2003kg}. More recently, the
QGP has also been described in terms of a condensate of $Z_3$ Wilson
lines \cite{Pisarski:2000eq} and by more refined quasiparticle models
based on the HTL-resummed entropy and (NLO) extensions thereof
\cite{Rebhan:2003wn}. These models have found support from
resummed perturbation theory \cite{Blaizot:2001nr} for temperatures $T\gsim 3
T_c$. However, they have difficulties explaining the dropping
of the thermal gluon screening mass in the vicinity of the phase
transition. An improved quasiparticle model \cite{Schneider:2001nf} shows the
correct temperature dependence of the Debye mass and reproduces lattice
thermodynamical quantities such as the pressure, the energy density
and the entropy density very well. The main new ingredient of this
model is a phenomenological parametrization of (de)confinement.\\[0.5cm]
In the present work, this improved quasiparticle model is extended to finite
quark chemical potential $\mu$. In section 2, a brief review of the
quasiparticle model with confinement is given. The extension of the
model to finite quark chemical potential $\mu$ is discussed in detail
in section 3. Numerical results are presented in chapter 4. The
phase boundary line $T_c(\mu)$ that separates the hadronic from the
QGP-phase is discussed and the quasiparticle model result is compared
to recent lattice simulations. Results for the pressure difference
from $\mu=0$ and the quark number density for various values of the chemical
potential $\mu$ are also presented and compared to recent lattice
simulations. A summary is given in section 5.
\section{Quasiparticle model with confinement}
It is possible to describe the EoS of hot QCD at vanishing quark chemical
potential $\mu$ to good approximation by the EoS of a gas of
quasiparticles with thermally generated masses, incorporating
confinement effectively by a temperature-dependent, reduced number of
thermodynamically active degrees of freedom. This method is briefly
outlined in this section. For a more detailed discussion the reader is
referred to ref.\cite{Schneider:2001nf}.\\[0.5cm]  
At very high temperatures, spectral functions for gluons or quarks of
the form $\delta(E^2-k^2-m^2(T))$ with $m(T) \sim g T$ are found in
HTL perturbative calculations. Here, $E$ is the particle energy,
$k$ the absolute value of its momentum, $m(T)$ its thermally generated
mass and $g$ the QCD coupling constant. As long as the spectral function
at lower temperatures  resembles qualitatively this asymptotic form, a
quasiparticle description is expected to be applicable. QCD dynamics
is then incorporated in the thermal masses of the quark and gluon
quasiparticles. These thermal masses are obtained from the self-energies
of the corresponding particles, evaluated at thermal momenta $k\sim
T$:

\begin{eqnarray}
&&m_q^2=m_{0q}^2+\frac{N_c^2-1}{8
N_c}\left(T^2+\frac{\mu^2}{\pi^2}\right)G^2(T,\mu)\label{eqm},\\
&&m_g^2=m_{0g}^2+\frac{1}{6}\left[\left(N_c+\frac{N_f}{2}\right)T^2+
\frac{3}{2\pi^2}\sum_q\mu_q^2\right]G^2(T,\mu)\label{egm}.
\end{eqnarray}

$N_f$ is the number of flavors, $N_c$ the number of colors. The
effective coupling strength $G$ is specified as

\begin{equation}
G(T,\mu=0)=\frac{g_0}{\sqrt{11 N_c-2 N_f}}
\left([1+\delta]-\frac{T_c}{T}\right)^\beta.\label{ecc}
\end{equation}

Setting $g_0=9.4$, $\beta=0.1$, the effective masses as given in
equations (\ref{eqm}) and (\ref{egm}), approach the HTL result at high
temperatures. (A small shift $\delta=10^{-6}$ helps fine-tuning at
$T\simeq T_c$). Because of the existence of
a heat bath background, new partonic excitations, plasmons
(longitudinal gluons) and plasminos (quark-hole excitations) are also
present in the plasma. However, their spectral strengths are
exponentially suppressed for hard momenta and large temperatures and
consequently these states are essentially unpopulated
\cite{LeBellac}.
The functional dependence of $m_g(T)$ on $T$ is based on the
conjecture that the phase transition is second order or weakly first
order which suggests an almost power-like behavior $m\sim
(T-T_c)^\beta$ with some critical exponent $\beta > 0$. It is assumed
that the pseudocritical form of the effective coupling constant given in
equation (\ref{ecc}) also provides the correct approximate expression
for the effective quark mass. This is supported by a non-perturbative
dispersion relation analysis for a thermal quark interacting with the
gluon condensate \cite{Schaefer:1998wd}.\\[0.5cm] 
Close to $T_c$ the picture of a non-interacting gas is not appropriate
because the driving force of the transition, the confinement process,
is not taken into account. Below $T_c$, the relevant degrees of
freedom are pions and other hadrons. Approaching $T_c$ from below,
deconfinement sets in and the quarks and gluons are liberated,
followed by a sudden increase in entropy and energy
density. Conversely, when approaching the phase transition from above,
the decrease in the thermodynamic quantities is not primarily caused
by increasing masses of the quasiparticles, but by the reduction of
the number of thermally active degrees of freedom due to the onset of
confinement. For example, gluons begin to form heavy clusters
(glueballs), so that the gluon density gets reduced as $T_c$ is
approached from above. This feature can be incorporated in the
quasiparticle picture by modifying the number of effective degrees of
freedom by a temperature dependent confinement factor $C(T)$: 

\begin{equation}
C(T,\mu=0)=C_0\left([1+\delta_c]-\frac{T_c}{T}\right)
\label{confinement_function}.
\end{equation}

The confinement factor is taken to be universal. The parameters $C_0$,
$\delta_c$ and $\beta_c$ are fixed by reproducing the entropy density
that results from lattice QCD thermodynamics. Since the results of
lattice calculations with dynamical quarks are still dependent on the
details of the simulations, $C_0$, $\delta_c$ and $\beta_c$ should be
finetuned for different lattice calculations.\\[0.5cm]
For homogeneous systems of large volume $V$, the Helmholtz free energy
$F$ is related to the pressure $p$ by $F(T,V)=-p(T)/V$. In the present
framework of a gas of quasiparticles, its explicit expression reads

\begin{equation}
p(T)=\frac{\nu_g}{6 \pi^2}\int_0^\infty\mbox{d}kC(T)f_B(E_k^g)\frac{k^4}{E_k^g}
+\frac{2 N_c}{3\pi^2}\sum_{i=1}^{N_f}\int_0^\infty\mbox{d}k
C(T)f_D(E_k^i)\frac{k^4}{E_k^i}-B(T).
\end{equation}

$\nu_g$ is the gluon degeneracy factor, $E_k^g=\sqrt{k^2+m_g^2(T)}$ is
the gluon energy, $E_k^q=\sqrt{k^2+m_q^2(T)}$ the quark energy,
$f_B(E_k^g)=(\exp((E_k^g)/T)-1)^{-1}$ the Bose-Einstein
distribution function of gluons and
$f_D(E_k^q)=(\exp((E_k^q)/T)+1)^{-1}$ the Fermi-Dirac
distribution function of quarks. The energy density $\epsilon$ and the
entropy density $s$ take the form 

\begin{equation}
\epsilon(T)=\frac{\nu_g}{2\pi^2}\int_0^\infty\mbox{d}kk^2 C(T)f_B(E_k^g)E_k^g
+\frac{2 N_c}{\pi^2}\sum_{i=1}^{N_f}\int_0^\infty\mbox{d}k k^2
C(T)f_D(E_k^i)E_k^i+B(T).
\end{equation}

and

\begin{equation}
s(T)=\frac{\nu_g}{2 \pi^2 T}\int_0^\infty\mbox{d}k k^2 C(T) f_B(E_k^g)
\frac{\frac{4}{3}k^2+m_g^2(T)}{E_k^g}
+\frac{2 N_c}{\pi^2 T}\sum_{i=1}^{N_f}\int_0^\infty\mbox{d}k k^2
C(T)f_D(E_k^i)\frac{\frac{4}{3}k^2+m_q^2(T)}{E_k^i}
\end{equation}

The function $B(T)$ is introduced to act as a
background field. It is necessary in order to maintain thermodynamic
consistency: $p$, $\epsilon$ and $s=\partial p/\partial T$ have to
satisfy the Gibbs-Duhem relation $\epsilon+p=Ts=T\partial p/\partial
T$. $B(T)$ basically compensates the additional $T$-derivatives from
the temperature-dependent masses in $p$ and thus is not an independent
quantity. Since $B(T)$ adds to the energy density of the
quasiparticles, it can be interpreted as the thermal vacuum energy
density. The entropy density, as a measure of phase space, is
unaffected by $B(T)$.

\section{Finite chemical potential}

The quasiparticle model reviewed in the previous section accurately
reproduces lattice thermodynamical quantities such as the pressure,
the energy density and the entropy density in the temperature range
$T_c < T\lsim 4 T_c$ at vanishing chemical potential 
\cite{Schneider:2001nf}. However, many
physical questions, e.g. the structure of quark cores in massive
neutron stars, the baryon contrast prior to cosmic confinement or the
evolution of the baryon number in the mid-rapidity region of central
heavy-ion collisions, require a detailed understanding of the EoS at
non-vanishing quark chemical potential. In this section, a
thermodynamically self-consistent extension of the quasiparticle model
to finite quark chemical potentials is presented. Results for various
observables are then computed and compared to finite $\mu$ lattice
results in the next section.\\[0.5cm]
At vanishing quark chemical potential, it is conjectured from
asymptotic freedom that QCD undergoes a phase transition from the
hadronic phase to the QGP phase.  At extremely high
density, cold quark matter is necessarily in the Color-Flavor-Locked
(CFL) phase in which quarks of all three colors and all three flavors
form cooper pairs. It is expected that this phase is separated from
the hadronic phase by the color superconducting 2SC phase. For a
review of the QCD phase diagram, the reader is referred to
\cite{Rajagopal:1999cp}. Our extension of the quasiparticle model provides a
straightforward way to map the EoS at finite temperature and vanishing
quark chemical potential into the $T-\mu$ plane without further
assumptions. However, since this continuous mapping relies on quark
and gluon quasiparticles, it cannot provide information about other
possible phases with a different (quasiparticle) structure. It is
therefore applicable in a limited range of not too large chemical
potentials.\\[0.5cm]
The pressure of an ideal gas of quark and gluon quasiparticles with
effective masses depending on temperature and quark chemical
potential, is given by

\begin{equation}
p(T,\mu)=\frac{\nu_g}{6
\pi^2}\int_0^\infty\mbox{d}kC(T,\mu)f_B(E_k^g)\frac{k^4}{E_k^g}
+\frac{N_c}{3\pi^2}\sum_{q=1}^{N_f}\int_0^\infty dk
C(T,\mu)[f_D^+(E_k^q)+f_D^-(E_k^q)]\frac{k^4}{E_k^q}-B(T,\mu),
\label{finite_mu_pressure}
\end{equation}

with $f_D^\pm(E_k^q)=(\exp((E_k^q\mp\mu)/T)+1)^{-1}$. The effective
coupling strength $G(T,\mu)$, the confinement factor $C(T,\mu)$ and
the mean field contribution $B(T,\mu)$ now also depend on the quark
chemical potential $\mu$. $B(T,\mu)$ is calculated in appendix
\ref{calc_B_T_mu}. The quark number density (which is related to the
baryon number density $n_B$ by $n_q=n_B/3$) retains the ideal gas form

\begin{equation}
n_q(T,\mu)=\frac{N_c}{\pi^2}\sum_{q=1}^{N_f}\int_0^\infty dk
C(T,\mu)[f_D^+(E_k^q)-f_D^-(E_k^q)]k^2,
\label{finite_mu_density}
\end{equation}

but with the confinement factor $C(T,\mu)$ included.\\[0.5cm]
In the previous section expressions for the coupling
$G(T,\mu=0)$ and the confinement factor $C(T,\mu=0)$ are
given. These expressions can be generalized to finite chemical
potential in a thermodynamically self-consistent way using Maxwell
relations. Imposing the Maxwell relation between the derivatives of
the quark number density and the entropy,

\begin{equation}
\left.\frac{\partial s}{\partial \mu}\right|_{T}
=\left.\frac{\partial n}{\partial T}\right|_{\mu}\quad\Longrightarrow\quad
\sum_i\left(\frac{\partial n}{\partial m_i^2}\frac{\partial
m_i^2}{\partial T}-\frac{\partial s}{\partial m_i^2}\frac{\partial
m_i^2}{\partial \mu}\right)=0\quad\mbox{and}\quad
\left(\frac{\partial n}{\partial C}\frac{\partial C}{\partial T}
-\frac{\partial s}{\partial C}\frac{\partial C}{\partial
\mu}\right)=0,
\end{equation}

yields a set of first oder quasilinear partial differential equations
for the effective coupling constant $G^2(T,\mu)$ and the confinement
factor $C(T,\mu)$:

\begin{eqnarray}
&&a_T(T,\mu;G^2)\frac{\partial G^2}{\partial T}
+a_\mu(T,\mu;G^2)\frac{\partial G^2}{\partial \mu}
=b(T,\mu;G^2),\label{flow_equation1}\\
&&c_T(T,\mu;G^2)\frac{\partial C}{\partial T}
+c_\mu(T,\mu;G^2)\frac{\partial C}{\partial \mu}
=0.\label{flow_equation2}
\end{eqnarray}

The coefficients $a_T$, $a_\mu$, $b$, $c_T$, $c_\mu$ depend on $T$,
$\mu$, $G^2$ but not on $C$. It can be solved by the method of
characteristics (see appendix \ref{moc}). The flow of the effective
coupling and the confinement factor is elliptic. In particular, one
finds 

\begin{equation}
a_T(T,\mu=0)=0,\quad a_\mu(T=0,\mu)=0,
\quad c_T(T,\mu=0)=0,\quad c_\mu(T=0,\mu)=0.
\end{equation}

\begin{figure}[p]
\begin{center}
\epsfig{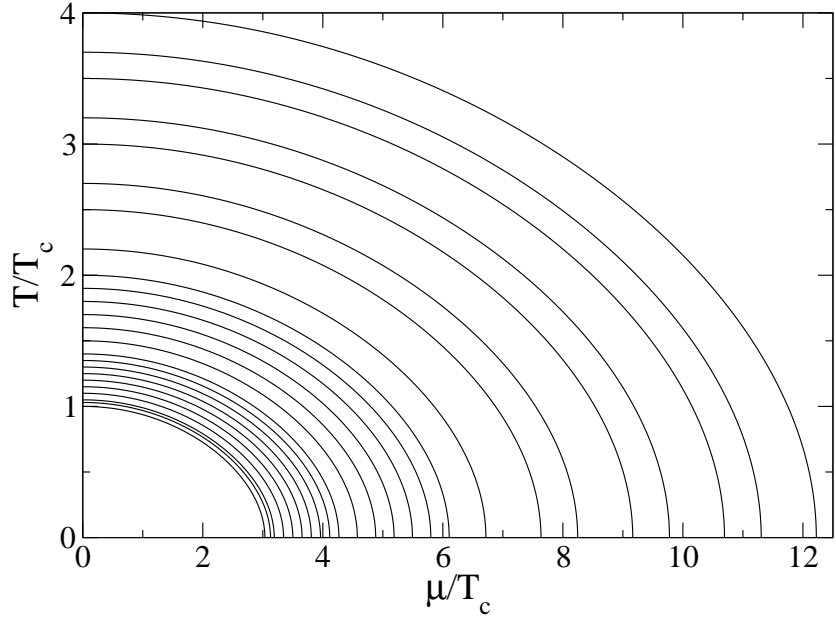}   
\caption{Characteristic curves of constant confinement factor
$C(T,\mu)=const$, obtained when solving eq.(\ref{flow_equation2}).}
\label{characteristics}
\end{center}
\end{figure}

\begin{figure}[p]
\begin{center}
\epsfig{bbllx=116,bblly=415,bburx=509,bbury=667,clip=,
 file=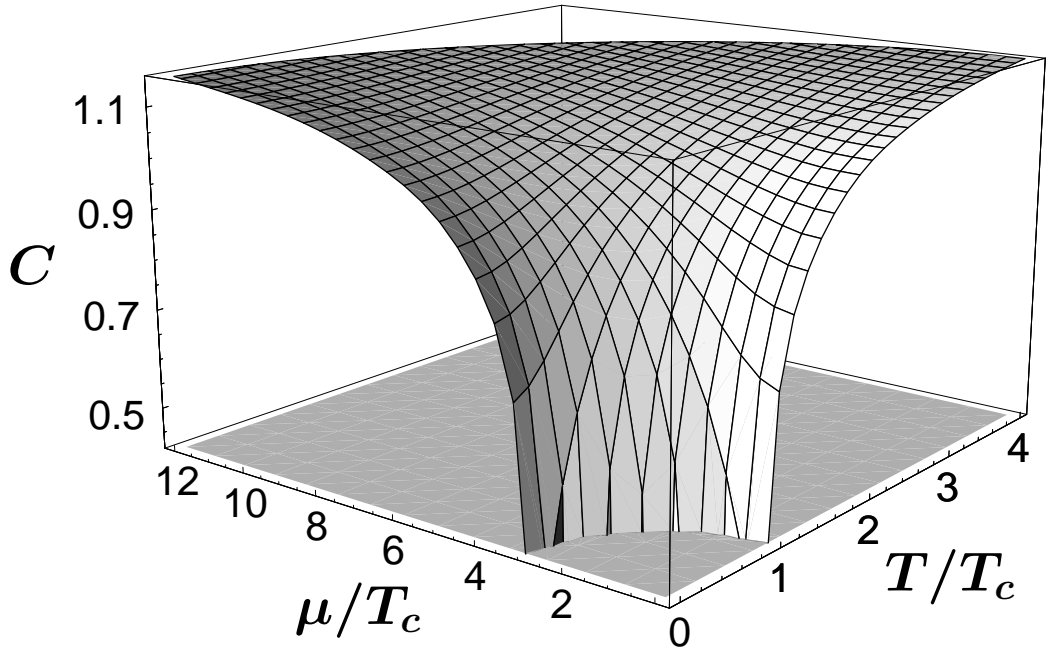,width=14cm}   
\caption{The confinement factor $C(T,\mu)$ as a function of the temperature $T$
and the quark chemical potential $\mu$.}
\label{confinement_factor}
\end{center}
\end{figure}

Therefore, the characteristics are perpendicular to both the $T$ and
the $\mu$ axis. This guarantees that specifying the coupling constant
and the confinement factor on the $T$ axis sets up a valid initial
condition problem. Plots of the characteristic curves and the
confinement factor are shown in figure \ref{characteristics} and
\ref{confinement_factor}.\\[0.5cm]

\section{Comparison with lattice results}

Simulations of QCD at finite chemical potential are extremely
difficult because the fermion determinant becomes complex. This
prohibits Monte Carlo importance sampling, which interprets the
measure as a probability factor and thus requires it to be
positive. While this problem remains unsolved, there are some
approaches which circumvent the sign problem and allow lattice
simulations for small chemical potentials $\mu\lsim T_c$. A review
comparing these methods in detail can be found in
\cite{Laermann:2003cv}.

\subsection{The phase boundary line}

In the case of vanishing chemical potential, universal arguments and
lattice simulations suggest a phase transition from the hadronic phase
to the QGP phase at a critical temperature $T_c$. For QCD with three
light flavors $m_u\sim m_d\sim m_s\sim 5$ MeV this transition is
expected to be first
order. For two light flavors $m_u\sim m_d\sim 5$ MeV and an infinitely
large $m_s$ there is no phase transition, only a smooth crossover
\cite{Fodor:2001pe}. This suggests there is a critical strange mass
$m_s^c$ at which one finds a second order phase transition. Lattice
calculations indicate that $m_s^c$ is about half of the physical mass
$m_s$. At finite quark chemical potential $\mu$ and vanishing $T$ a
first order phase transition is predicted. For the physical $m_s$ this
implies that there is a first order phase transition for small $T$
and large $\mu$ which ends at a critical point $(T^*,\mu^*)$. At this
point the phase transition is of second order. For large $T$ and small
$\mu$ the two phases are separated by a crossover. We refer to the
line $T_c(\mu)$ that separates the hadronic phase from the QGP phase
as the ``phase boundary line''. In the literature
\cite{D'Elia:2002gd,deForcrand:2002ci,deForcrand:2003hx} this line is 
also frequently called the ``pseudocritical line''. $T_c(\mu)$ 
has been calculated on the lattice for 
$N_f=4$ \cite{Fodor:2001au,D'Elia:2002gd},
$N_f=2$ \cite{deForcrand:2002ci} and $N_f=3$ \cite{deForcrand:2003hx}
flavors of quarks up to quark chemical potentials $\mu$ of order
$T_c$. In the following we focus on the three-flavor results where the
critical line has been calculated with an accuracy up to terms of
order $(\mu/T)^6$. There, a Wilson gauge action and three degenerate
flavors of staggered quarks have been employed, with bare
masses in the range $0.025<am<0.04$, where $a$ denotes the lattice
spacing. The finite volume scaling behavior was monitored by using
three lattice sizes, $8^3\times 4$, $10^3\times 4$ and $12^3\times
4$.\\[0.5cm] 
In our quasiparticle model, the sudden decrease of the pressure, the
energy density, the quark number density and the entropy density
caused by gluons and quarks getting trapped in glueballs and hadrons 
when $T_c$ is approached from above, is parametrized by
the confinement factor $C(T,\mu)$. Consequently, it is natural to
relate the critical line to the characteristic curve of the confinement
factor through $T_c(\mu)$, as long as $\mu$ is small and the nature of
the quasiparticles does not change qualitatively.\\[0.5cm]
In order to calculate the confinement factor at finite chemical potential, we
need to specify a valid initial condition, e.g. $C(T,\mu=0)$. The
functional form of $C(T,\mu=0)$ is set by
eq.(\ref{confinement_function}). We have employed the following set of
parameters, as found in ref.\cite{Schneider:2001nf}:

\begin{center}
\begin{tabular}{|c|r|r|r|}\hline
   & $C_0$ & $\delta_c$ & $\beta_c$ \\ \hline
3 flavors & 1.03 & 0.02 & 0.2\\ \hline
\end{tabular}
\end{center}

\begin{figure}[ht!]
\begin{center}
\epsfig{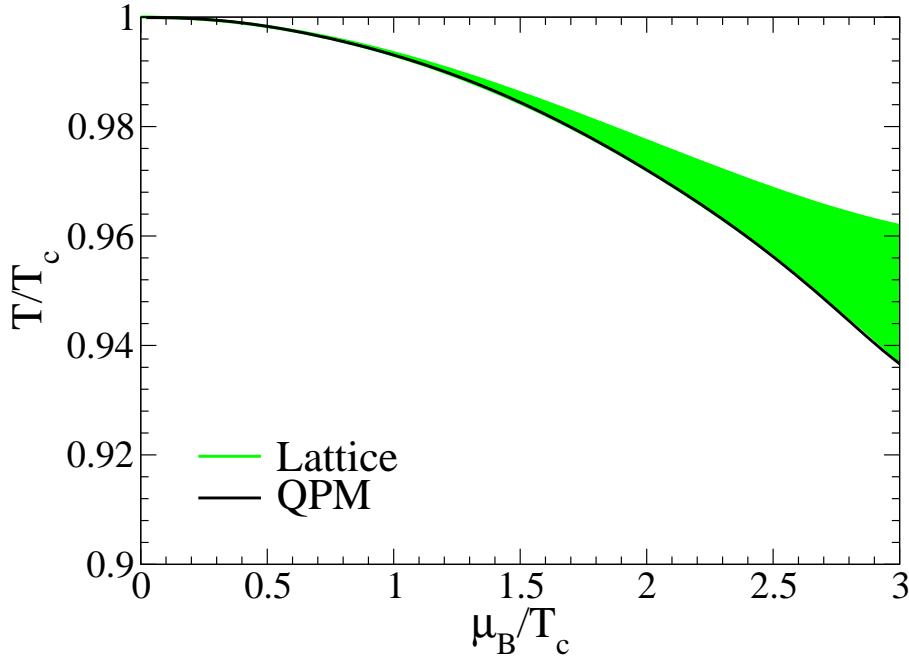}   
\caption{The phase boundary line $T_c(\mu)$ calculated with the
quasiparticle model for $N_f=3$.  The shaded band shows the one-sigma
error band obtained in lattice calculations in \cite{deForcrand:2003hx}.}
\label{pseudocritical_line}
\end{center}
\end{figure}

We have checked that the form of the phase boundary line in the
quasiparticle model depends only weakly on the exact choice of
parameters and a small difference only shows up for values much larger
than the range of $\mu$ covered by the lattice simulations. The
lattice phase boundary line and our result is shown in figure
\ref{pseudocritical_line}.
The quasiparticle result is within the lattice estimate for
$\mu_B\lsim2.5 T_c$ and deviates only slightly from the lattice result
for larger chemical potentials.

\subsection{Thermodynamical quantities}

There have been lattice calculations of thermodynamical quantities at
finite chemical potential for $N_f=2+1$ \cite{Fodor:2002km} and
$N_f=2$ \cite{Allton:2003vx} flavors of quarks. 
In the following we focus on results from \cite{Allton:2003vx} where a
p4-improved staggered action on a $16^3\times 4$ lattice was
used. There, the $N_\tau$ dependence is known to be small, in contrast
to standard staggered fermion actions which show substantially larger cut-off
effects. Estimates of the pressure, the quark number density and
associated susceptibilities as functions of the quark chemical
potential were made via a Taylor series expansion of the thermodynamic
grand canonical potential $\Omega$ up to fourth order.\\[0.5cm]
To calculate thermodynamical quantities within the quasiparticle model,
we need to fix the parameters of the effective coupling constant and
the confinement factor. Our calculations have shown that the
results are not sensitive to the detailed choice of parameters for the
effective coupling $G$. We have therefore used the parameters from
ref.\cite{Schneider:2001nf} in our calculations. In principle, the
parameters of the confinement factor can be fixed by comparing our
calculations to lattice results at vanishing chemical
potential. However, in ref.\cite{Allton:2003vx} no $\mu=0$ lattice
data is given. Since lattice calculations including
quarks give slightly different results depending on which action has
been used, fitting the parameters in $C(T,\mu=0)$ by comparing
quasiparticle results to lattice data from a different simulation is
not feasible and would lead to large differences. Consequently, we
directly used the finite $\mu$ lattice results for fitting. Good
agreement with the lattice thermodynamical observables was found for
the following sets of parameters: 

\begin{center}
\begin{tabular}{|c|r|r|r|}\hline
   & $C_0$ & $\delta_c$ & $\beta_c$ \\ \hline
Set A & 1.05 & -0.016 & 0.15\\ \hline
Set B & 1.12 & 0.02 & 0.2\\ \hline
\end{tabular}
\end{center}

\begin{figure}[p]
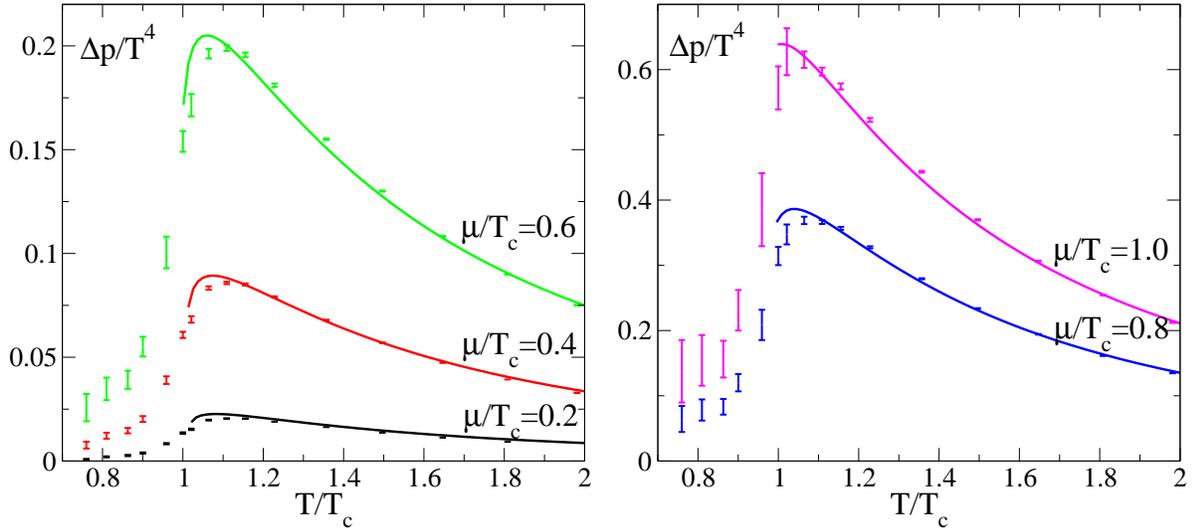

\begin{center}
\epsfig{bbllx=40,bblly=27,bburx=590,bbury=523,clip=,
 file=pd1_set_A.eps,width=7.8cm}   
\epsfig{bbllx=40,bblly=27,bburx=590,bbury=523,clip=,
 file=pd2_set_A.eps,width=7.8cm}   
\caption{The normalized pressure difference $\Delta p(T,\mu)/T^4$ as a
function of temperature compared to lattice results from
\cite{Allton:2003vx} (symbols).}
\label{pressure_discrepancy}
\end{center}
\end{figure}

\begin{figure}[p]
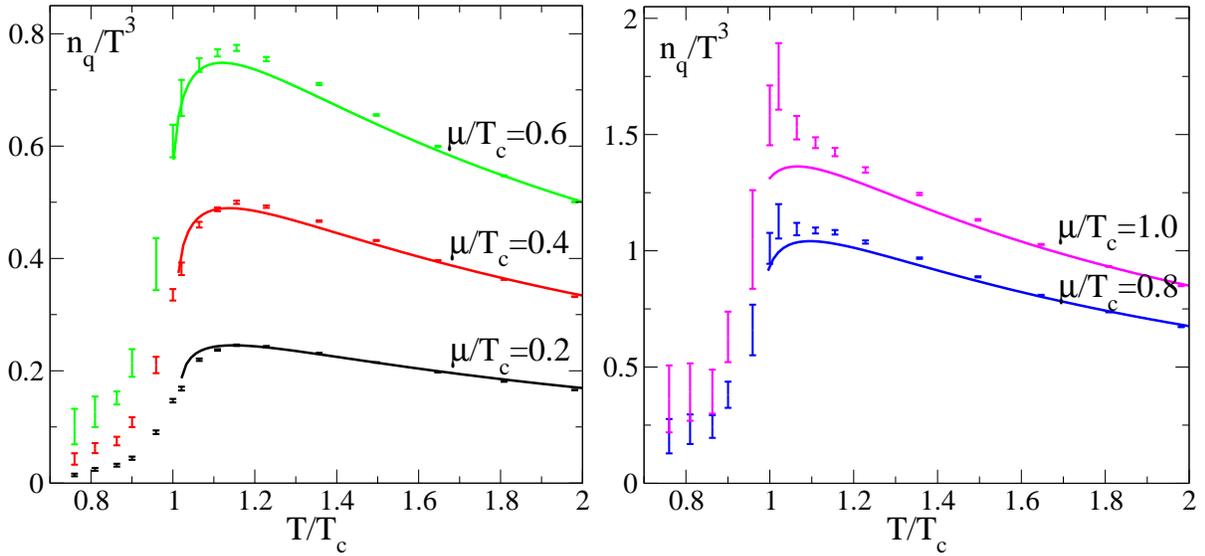

\begin{center}
\epsfig{bbllx=64,bblly=27,bburx=590,bbury=523,clip=,
 file=qnd1_set_A.eps,width=7.8cm}   
\epsfig{bbllx=53,bblly=27,bburx=590,bbury=523,clip=,
 file=qnd2_set_A.eps,width=7.95cm}
\caption{The normalized quark number density $n_q(T,\mu)/T^3$ as a
function of temperature compared to lattice results from
\cite{Allton:2003vx} (symbols).}
\label{quark_number_density_plot1}
\end{center}
\end{figure}

While set A reproduces the lattice thermodynamical results
slightly better, set B is in better agreement with the parameters
found in \cite{Schneider:2001nf} for $\mu=0$ lattice
simulations.\\[0.5cm]
The temperature dependence of the normalized pressure difference
$\Delta p(T,\mu)=(p(T,\mu)-p(T,\mu=0))/T^4$ is shown in figure
\ref{pressure_discrepancy} and that of the normalized quark number
density $n_q(T,\mu)/T^3$ in figure \ref{quark_number_density_plot1}. 
Whereas the computation of the quark number density from equation
(\ref{finite_mu_density}) is straightforward, a numerical evaluation
of (\ref{finite_mu_pressure}) is difficult because of the derivatives
of the effective masses and the confinement factor in $B(T,\mu)$ (see
expressions in appendix \ref{calc_B_T_mu}). It turns out that it is
simpler to calculate the pressure difference using the
following relation:

\begin{equation}
\Delta p(T,\mu)=\frac{1}{T^4}\int_0^\mu\mbox{d}\mu'n_q(T,\mu').
\end{equation}

The lattice pressure difference is well reproduced even for the
largest values of the chemical potential. The quark number density is
in very good agreement with the lattice data for $\mu/T_c=0.2$ and
$0.4$. For larger values of $\mu$ our calculations underestimate the
magnitude of the lattice results close to $T_c$, but show the same
qualitative features.  

\section{Summary}
We have presented a quasiparticle description of the QCD EoS at finite
temperature and quark chemical potential. 
Our main modification as compared to previous
work is the inclusion of finite quark chemical potential in a
thermodynamically consistent way. We have first reviewed our 
improved quasiparticle model which schematically includes
confinement. We have shown how Maxwell relations can be used to
construct the effective coupling $G(T,\mu)$ and the
confinement factor $C(T,\mu)$ at finite chemical potential. We then
used this model to calculate the phase boundary line $T_c(\mu)$ and
the normalized pressure difference $\Delta
p(T,\mu)=(p(T,\mu)-p(T,\mu=0))/T^4$ and the normalized quark number
density $n_q(T,\mu)/T^3$. We compared our results to recent lattice
calculations and found remarkably good agreement even for large quark
chemical potentials $\mu\sim T_c$. 

\appendix

\section{Calculation of $B(T,\mu)$\label{calc_B_T_mu}}

The ``background field'' quantity $B(T,\mu)$ appearing in
eq.(\ref{finite_mu_pressure}) can be obtained from the Gibbs-Duhem
relation  
\begin{equation}
\epsilon+p=Ts+\mu n=T\frac{\partial p}{\partial T}
+\mu\frac{\partial p}{\partial\mu}.
\end{equation}
The left hand side reads:
\begin{equation}
\epsilon+p=\frac{N_c N_f}{3 \pi^2}\int_0^\infty dk [f_D^+ + f_D^-]
C(T,\mu) k^2\left( \frac{4 k^2 + 3 m_q^2}{E_k}\right)\label{finitemu1}.
\end{equation}
To evaluate the right-hand side, derivatives of $f_D^\pm(E_k^q)$ with
respect to $T$ and $\mu$ are rewritten as derivatives with respect to
$k$. After an integration by parts, the first term on the right-hand
side reads
\begin{eqnarray}
T \frac{\partial p}{\partial T} &=& T\frac{N_c N_f}{3 \pi^2} \int_0^\infty
dk f_D^+\left(\frac{\partial C}{\partial T}\frac{k^4}{E_k}-C(T,\mu)\frac{3 
  k^2}{2 E_k}\frac{\partial m_q^2}{\partial T} + C(T,\mu)(E_k+\mu) +
C(T,\mu)\frac{k^4}{T E_k}\right)\nonumber\\
 &+& T\frac{N_c N_f}{3 \pi^2} \int_0^\infty
dk f_D^-
\left(\frac{\partial C}{\partial T}\frac{k^4}{E_k}-C(T,\mu)\frac{3 
  k^2}{2 E_k}\frac{\partial m_q^2}{\partial T} + C(T,\mu)(E_k-\mu) +
C(T,\mu)\frac{k^4}{T E_k}\right)\nonumber\\
&-& T\frac{\partial B(T,\mu)}{\partial T},\label{finitemu5}
\end{eqnarray}
and the second term is given by
\begin{eqnarray}
\mu \frac{\partial p}{\partial \mu}  &=& \mu\frac{N_c N_f}{3
   \pi^2}\int_0^\infty dk f_D^+\left(\frac{\partial C(T,\mu)}{\partial
   \mu} \frac{k^4}{E_k}-C(T,\mu) \frac{\partial m_q^2}{\partial
   \mu}\frac{3 k^2}{2 E_k}-C(T,\mu) 3 k^2\right)\nonumber\\
&+& \mu\frac{N_c N_f}{3
   \pi^2}\int_0^\infty dk f_D^-\left(\frac{\partial C(T,\mu)}{\partial
   \mu} \frac{k^4}{E_k}-C(T,\mu) \frac{\partial m_q^2}{\partial
   \mu}\frac{3 k^2}{2 E_k}+C(T,\mu) 3 k^2\right)
-\mu\frac{\partial B(T,\mu)}{\partial \mu}.
\label{finitemu9}
\end{eqnarray}
Substituting (\ref{finitemu1}), (\ref{finitemu5}) and
(\ref{finitemu9}) in the Gibbs-Duhem relation yields a partial
differential equation of the type
\begin{equation}
x\frac{\partial f(x,y)}{\partial x}+y\frac{\partial f(x,y)}{\partial
y}=\mathcal{I}(x,y).
\end{equation}
It has the general solution
\begin{equation}
f(x,y)=\int^xdt
\mathcal{I}(t,\frac{y}{x}t)+\mathcal{H}\left(\frac{y}{x}\right).
\end{equation}
Here, $\mathcal{H}(y/x)$ is a solution of the homogeneous
equation. Returning to our case, $\mathcal{H}(\mu/T)$ becomes an
arbitrary function of the ratio $\mu/T$ to be fixed by
boundary conditions. For $\mu\to 0$, $\mathcal{H}(\mu/T)$ does not
depend on $T$ anymore and therefore has to be identified with an
integration constant $B_0$. Provided that $\mathcal{H}(\mu/T)$ is a
continuous function it must be close to $B_0$ for small $\mu/T$.  The
first term in a Taylor expansion of $\mathcal{H}(\mu/T)$ vanishes and
the series starts only at order $(\mu/T)^2$. Therefore we identify 
$\mathcal{H}(\mu/T)$ with the constant $B_0$ for all $\mu$ under
consideration. Assembling all pieces, the final result reads
\begin{eqnarray}
B(T,\mu)&=&B_1(T,\mu)+B_2(T,\mu)+B_0,\quad\nonumber\\
B_1(T,\mu)&=&\frac{N_c N_f}{3 \pi^2}\int_0^\infty dk
\int_{T_c}^T d\tau\left[f_D^+(E_k^q)+
f_D^-(E_k^q)\right]\left(\frac{\partial C}{\partial \tau}
+\frac{\mu}{T}\frac{\partial C}{\partial \left( \frac{\mu}{T}\tau
  \right)}\right)\frac{k^4}{E_k^q},\nonumber\\
B_2(T,\mu)&=&- \frac{N_c N_f}{2 \pi^2} \int_0^\infty dk
\int_{T_c}^T d\tau \,C\left[f_D^+(E_k^q)+
 f_D^-(E_k^q)\right]\left(\frac{\partial
   m_q^2}{\partial \tau}
+\frac{\mu}{T}\frac{\partial
   m_q^2}{\partial \left( \frac{\mu}{T}\tau
  \right)}\right)\frac{k^2}{E_k^q},
\label{finitemu11}
\end{eqnarray}
where the explicit $\tau$-dependence in $C(\tau,\mu/T\,\tau)$,
$m_q(\tau,\mu/T\,\tau)$ and  $E_k^q(\tau,\mu/T\,\tau)$ has been
suppressed for the sake of lucidity.

\section{Method of characteristics\label{moc}}

Equations (\ref{flow_equation1}) and (\ref{flow_equation2}) are a set
of coupled quasilinear first order partial differential equations for
the effective coupling constant $G^2(T,\mu)$ and the confinement
factor $C(T,\mu)$. Equation (\ref{flow_equation1}) does not depend on
$C(T,\mu)$. Thus we can first solve this equation for $G^2(T,\mu)$ and
insert the result in equation (\ref{flow_equation2}).\\[0.5cm]
The usual method found in textbooks is to reduce a quasilinear
partial differential equation of the form

\begin{equation}
a_T(T,\mu;X)\frac{\partial X}{\partial T}+a_\mu(T,\mu;X)
\frac{\partial X}{\partial\mu}=c(T,\mu;X)\label{flow_equation3}
\end{equation}

to  a system of coupled ordinary differential equations,

\begin{equation}
\frac{dT(s)}{ds}=a_T,\quad
\frac{d\mu(s)}{ds}=a_\mu,\quad
\frac{dX(s)}{ds}=c.
\end{equation}

This determines the characteristic curves $T(s)$, $\mu(s)$, and the
evolution of $X$ along such a curve, given an initial value. However,
this method is not well suited for numerical use which is necessary for
non-trivial $a_T$, $a_\mu$ and $c$. Rewriting equation
(\ref{flow_equation3}) as 

\begin{equation}
a_T\left(\frac{dX}{dT}-\frac{\partial X}{\partial \mu}\frac{d\mu}{d T}\right)
+a_\mu\frac{\partial X}{\partial \mu}=c
\Longrightarrow \frac{\partial
X}{\partial\mu}\left(\frac{a_\mu}{a_T}dT-d\mu\right)
=\frac{c}{a_T}dT-dX,
\end{equation}

we find the equation $a_\mu dT-a_T d\mu=0$ for the characteristics and
$c dT-a_TdX=0$ for the evolution of $X$. These equations can easily be
solved numerically.

\end{document}